\documentclass{article}
\usepackage{spconf,amsmath,epsfig}

\begin{document}\sloppy

\def\x{{\mathbf x}}
\def\L{{\cal L}}

\title{Hybrid Point Cloud Attribute Compression Using Slice-based Layered Structure and Block-based Intra Prediction}
%
\name{Yiting Shao$[1]$, Qi Zhang$[1]$, Ge Li$[1]$, Zhu Li$[2]$}
\address{Peking University Shenzhen Graduate School$[1]$, UMKC$[2]$}

\maketitle

\begin{abstract}
Point cloud compression is a key enabler for the emerging applications of immersive visual communication, autonomous driving and smart cities, etc. In this paper, we propose a hybrid point cloud attribute compression scheme built on an original layered data structure. First, a slice-partition scheme and geometry-adaptive k dimensional-tree (kd-tree) method are devised to generate the four-layer structure. Second, we introduce an efficient block-based intra prediction scheme containing a DC prediction mode and several angular modes, in order to exploit the spatial correlation between adjacent points. Third, an adaptive transform scheme based on Graph Fourier Transform (GFT) is Lagrangian optimized to achieve better transform efficiency. The Lagrange multiplier is off-line derived based on the statistics of color attribute coding. Last but not least, multiple reordering scan modes are dedicated to improve coding efficiency for entropy coding. In intra-frame compression of point cloud color attributes, results demonstrate that our method performs better than the state-of-the-art region-adaptive hierarchical transform (RAHT) system, and on average a 29.37$\%$ BD-rate gain is achieved. Comparing with the test model for category 1 (TMC1) anchor's coding results, which were recently published by MPEG-3DG group on 121st meeting, a 16.37$\%$ BD-rate gain is obtained. 
\end{abstract}
\section{Introduction}
Point clouds are being widely applied in many fields such as cultural heritage, immersive communication, autonomous navigation, etc \cite{usecase}. Immersive visual communication, such as virtual reality (VR) and augmented reality (AR) applications, is showing fresh new capabilities and experiences and gaining momentum in the industry \cite{wu}. The core to achieving this is constructing an effective and efficient 3D scene capture, compression and communication system \cite{lin}. Autonomous navigation and cultural heritage reconstruction applications also require similar capabilities and even higher resolution and fidelity of point clouds.

Point Cloud Compression is established as a working group under MPEG to develop novel solutions to compress 3D geometry and attribute information. This is the key enabler as well as a bottleneck in achieving the immersive communication and autonomous driving vision. 

The traditional method to generate 3D models from point clouds is using triangle or polygon meshes to reconstruct the underlying surface model \cite{mesh}. The method needs to estimate the connectivity among points, which causes high complexity and may introduce artifacts. As 3D rendering technologies and computing machines are developing rapidly, producing 3D models with amounts of discrete points becomes more practicable \cite{high}. 3D point clouds have gradually been favored over meshes for representing the surfaces of 3D objects and scenes.

A Point cloud usually contains millions of points and each point is associated with geometry positions and attribute information. Point clouds are irregularly distributed in 3D space without structured points arranged on a grid, where traditional 2D picture coding tools and video coding schemes cannot effectively work. A challenge in 3D point cloud compression is figuring out how to exploit the spatial and temporal correlation among massive discrete points, which are associated with 3D positions and attributes. 

There are three main types of compression in point clouds: geometry compression, attribute compression and dynamic motion-compensated compression. Geometry compression aims to code 3D point coordinates in point clouds. Attribute compression intends to reduce the redundancy among point cloud attributes. Dynamic motion-compensated compression targets dynamic point cloud sequences compression. In this paper, we focus on intra-frame compression of point cloud color attributes.

\begin{figure*}[!t]
	\centering
	\includegraphics[width=1\textwidth]{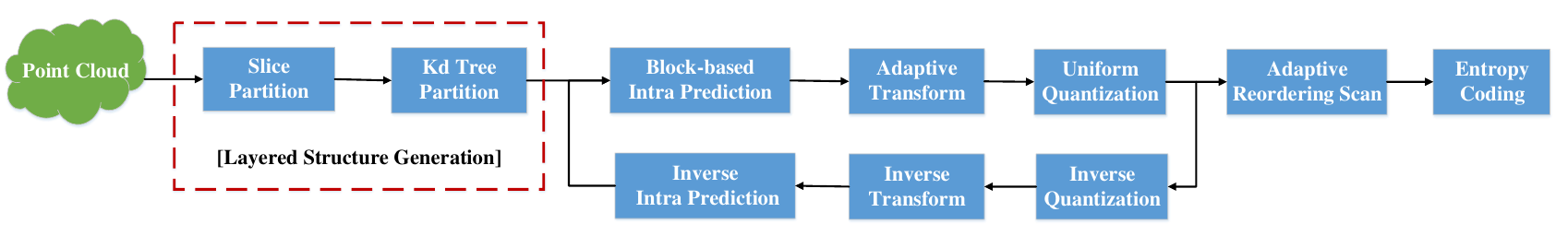}
	\caption{Schematic flow of the proposed point cloud attribute compression procedures.}
	\label{fig:overview}
\end{figure*}

Cha's GFT-based scheme \cite{zhang}, Rufael's 2D-mapping method \cite{design} and Ricardo's RAHT scheme \cite{ricardo} were three typical works on point cloud attribute compression. These contributions were focusing on the improvement of transform techniques. Cha firstly introduced GFT to code point cloud attributes and achieved better coding performance than traditional octree-based methods. But the framework of GFT method was not well optimized. In \cite{design}, Rufael projected point cloud color attributes to 2D grids and used JPEG codec to compress grids. The method was efficient but may introduce artifacts because of the 3D-to-2D mapping process. The state-of-the-art was Ricardo's work in \cite{ricardo}. They devised an original hierarchical sub-band transform which was more computationally efficient than GFT scheme, while GFT performed better than RAHT on attribute compression.

Actually, the compression performance of a point cloud attribute encoder depends on many coding components. In this paper, we propose an innovative hybrid point cloud attribute coding scheme. The scheme is embodied in layered structure generation, block-based intra prediction, adaptive GFT-based transform, optimized reordering scan before entropy coding. First, a slice-partitioning scheme and a block-division method are adopted to generate the layered structure. Second, on the layered structure, we introduce an efficient block-based intra prediction scheme, which provides a DC mode and five angular modes. The sum of absolute transformed difference (SATD) is used to choose the best mode. Later, Lagrangian optimized GFT and DCT two transform modes are adaptively adopted to achieve better transform efficiency for different types of point clouds. The Lagrange multiplier is off-line derived based on the statistics of color attribute coding. Before entropy coding, multiple reordering scan modes are dedicated to improve coding efficiency. For additional point cloud attributes, such as the normal information, this coding framework can also perform effectively and efficiently.

The rest of this paper is organized as follows. Section 2 presents related works. An overview of the proposed coding scheme is shown in Section 3. Point cloud layered structure, block-based intra prediction, adaptive GFT-based transform and the reordering scan scheme are detailed in Section 4, 5, 6 and 7, respectively. In Section 8, we present experimental results to evaluate the proposed scheme. Finally, we conclude in Section 9.

\begin{figure}[t]
	\centering
	\includegraphics[width=0.4\textwidth]{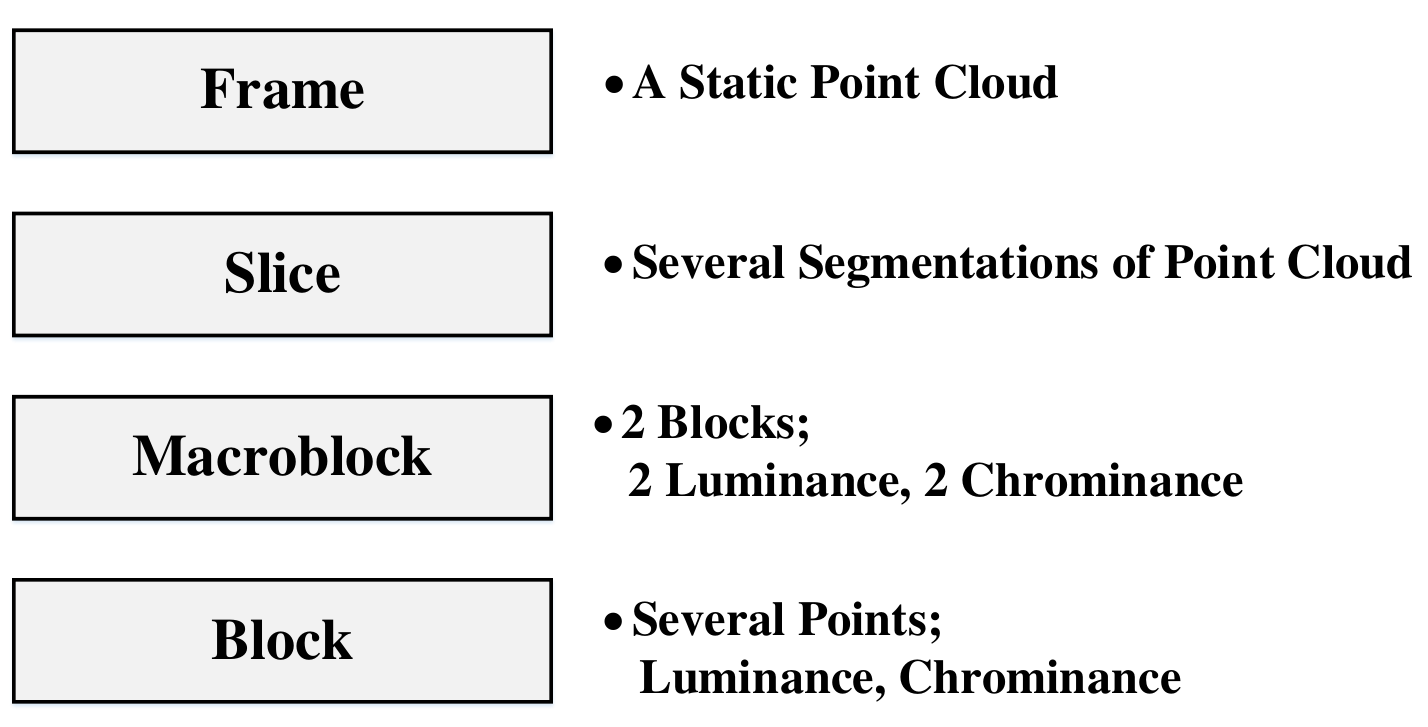}
	\caption{Point cloud layered data structure.}
	\label{fig:layer}	
\end{figure}

\section{Related Works}

There are some works already on point cloud attribute compression. In order to deal with the massive discrete points, the first step is to set up a regular structure for point clouds. Schnabel in \cite{soctree} first applied the octree structure in the static point cloud single-rate compression. Later, the octree method was further developed for progressive point cloud compression in \cite{huang} and dynamic point cloud coding in \cite{real}\cite{design}. K-D tree was another popular method to represent point cloud. Devillers in \cite{kdtree} adopted the kd-tree approach to recursively subdivide the bounding box of a point cloud. Shao in \cite{shao} devised an improved kd-tree scheme to uniform partition without empty blocks. There were other representations of point cloud. Merry in \cite{merry} built a minimum spanning tree for point cloud sing-rate compression, but it performed very poorly on models with disjoint components. Fan in \cite{fan} tried to construct the level of details (LOD) hierarchy through an iterative point clustering process, and Anis in \cite{anis} modeled the point cloud based on consistently-evolving subdivisional triangular meshes. 

\begin{figure}[t]
	\centering
	\includegraphics[width=0.45\textwidth]{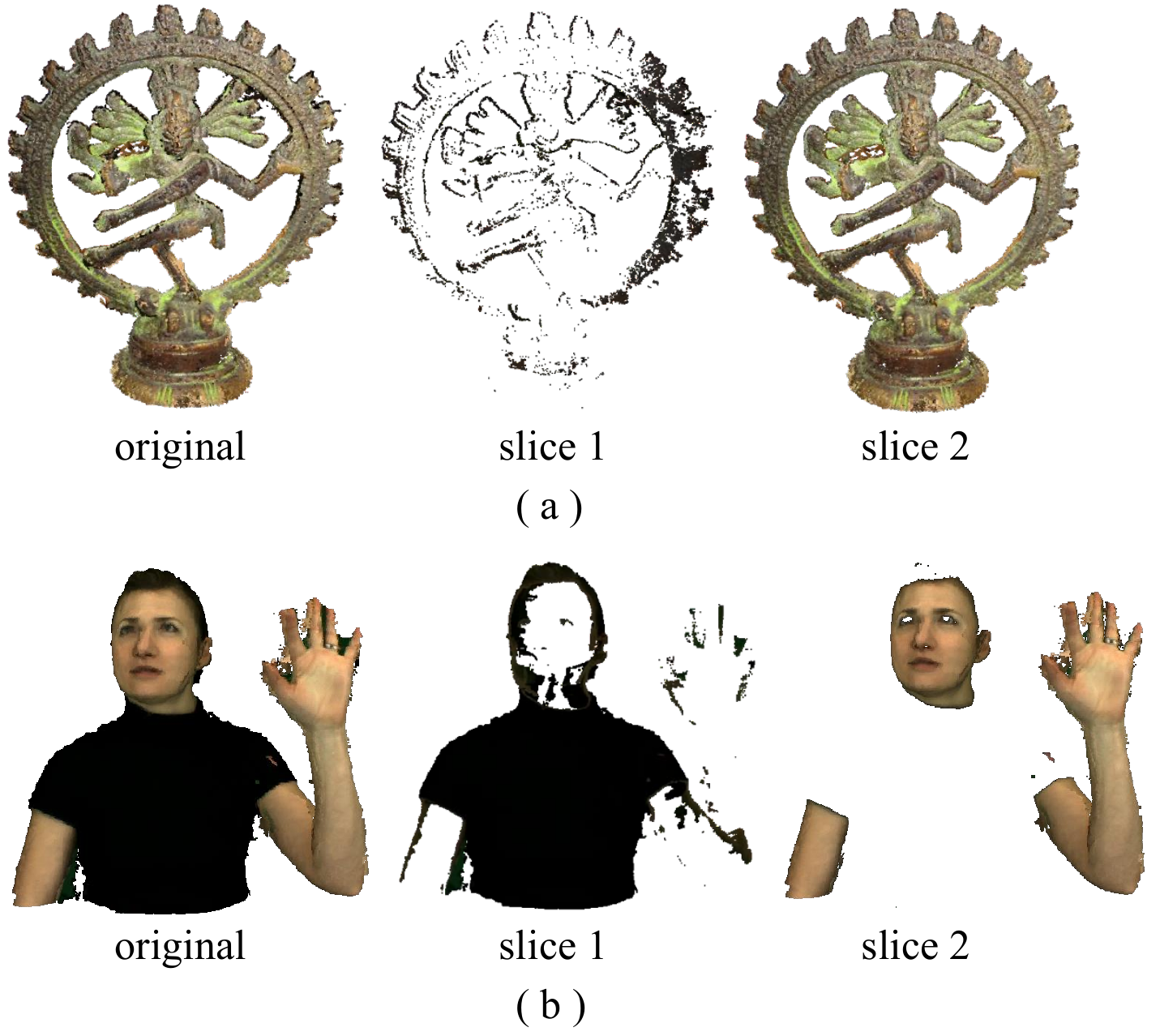}
	\caption{Examples on slice partition: (a) Shiva-35-vox12.ply, (b) Sarah.ply.}
	\label{fig:slice}	
\end{figure}

Current intra prediction schemes on attribute compression did not work well. Inspired by traditional hybrid video coding structures such as H.264/AVC \cite{AVC} and HEVC \cite{HEVC}, Robert et al in \cite{cohen} proposed the 3D intra prediction method with the octree partition. They projected the reconstructed attributes in neighboring blocks onto the adjacent edge planes of the current block and adopted the projection values as references. The prediction performance depended on the blocks after octree partition, since the number of points in blocks was usually different. Merry in \cite{merry} predicted future vertices using a linear predictor on the basis of a spanning tree, which was resource intensive to generate. 

Many improvements on attribute compression were based on transform. Rufael in \cite{design} mapped color attributes to a 2D grid based on a deep first octree traversal and used DCT-based JPEG codec to encode point cloud colors. The transform scheme made some progress in point cloud attribute compression, but it introduced blocking effects. GFT was first proposed in \cite{zhang}. Zhang et al formed a graph in each octree leafnode by using edges to connect nearby occupied voxels limited to one unit apart. Then GFT on the graph was used to encode point cloud attributes. Results showed that GFT had better coding performance than DCT, but there were still some problems need to be solved, such as sub-graph problems. Robert in \cite{bob} adopted k-nearest-neighbor (KNN) method to connect more distant points in a graph; nevertheless, experiments showed that the KNN method cannot solve sub-graph problems thoroughly. A compression framework that combining kd-tree structure and Laplacian sparsity optimized GFT was proposed in \cite{shao}. It showed that optimized GFT achieved better performance than general GFT, but it would not run well for all types of point cloud datasets. The RAHT-based method in \cite{ricardo} is the state-of-the-art for point cloud intra-frame compression. Ricardo et al devised a hierarchical sub-band transform and Laplace distribution-assumed arithmetic coding.

MPEG-3DG Ad Hoc Point Cloud Coding (PCC) Group is focusing on developing point cloud compression standards and has made much progress on static point cloud compression, dynamic point cloud compression, and dynamic acquisition point cloud compression. The latest test models for three categories released by MPEG in \cite{cfp} were RAHT-based compression scheme from 8i, video-based coding framework published by Apple and hierarchical coding tools also by Apple, respectively.

\section{Overview of Proposed Attribute Coding Scheme}

\begin{figure}[t]
	\centering
	\includegraphics[width=0.4\textwidth]{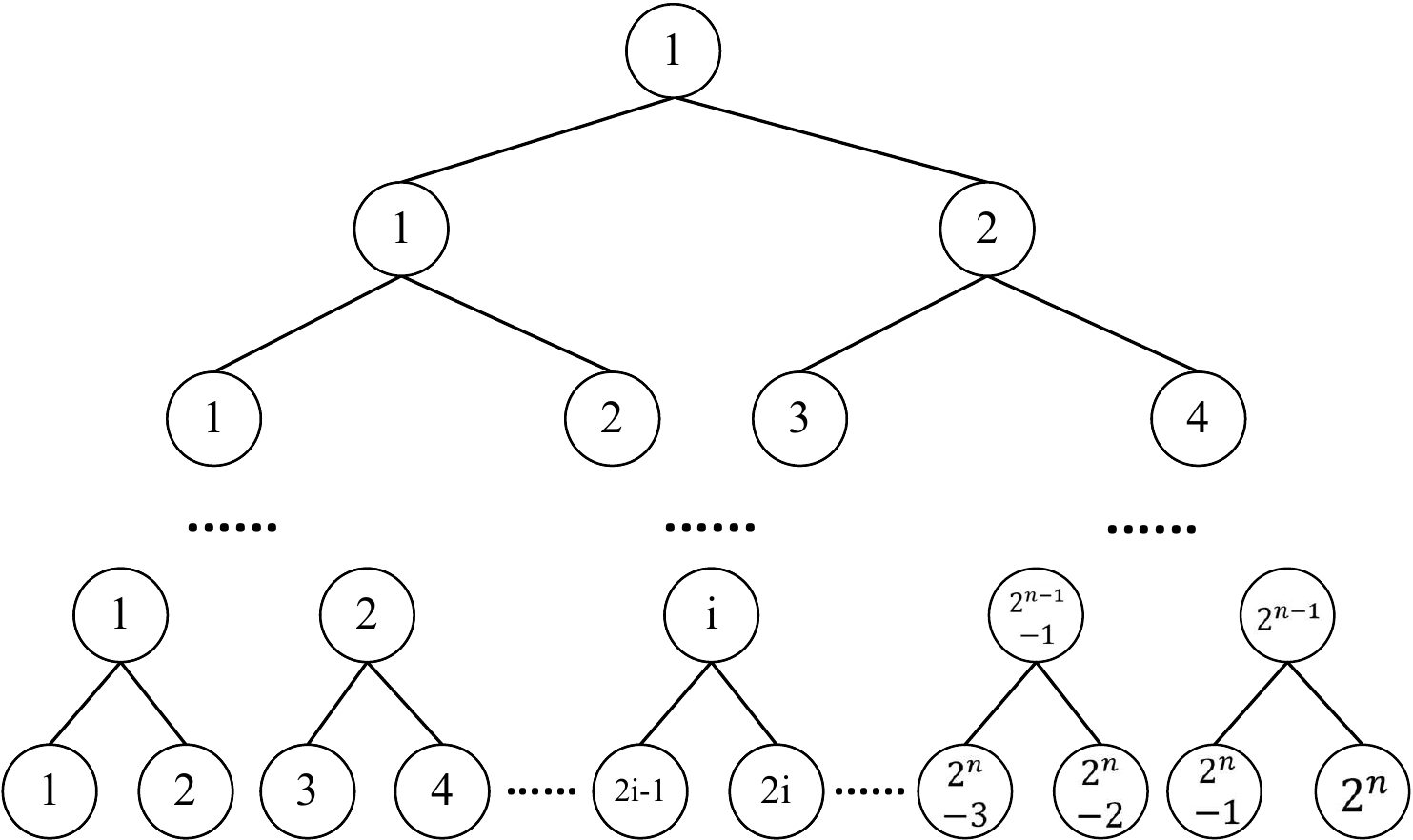}
	\caption{Kd-tree partition and the index of blocks.}
	\label{fig:kdtree}	
\end{figure}

\begin{table}[t]
	\caption{Kd-tree details for two example datasets} \label{intra}
	\label{tab:tree}
	\renewcommand\arraystretch{1.4}
	\begin{tabular}{|c|c|c|c|c|}
		\hline
		Frame Name & All\_point & Depth & Points & Blocks\\
		\hline
		Shiva\_35 & 1009132 & 13 & 123/124 & 8192 \\
		Sarah & 301626 & 11 & 147/148 & 2048 \\
		\hline
	\end{tabular}
\end{table}

We assume that geometry has been coded via a separate pipeline and geometry decoder would pass decoded geometry as side information to the attribute encoder. Without loss of generality, we use color attributes as an example of point cloud attributes in this paper. 

A schematic overview of the complete point cloud attribute compression procedures is illustrated in Figure \ref{fig:overview}. The proposed mechanism is mainly embodied in the layered structure generation, block-based intra prediction, adaptive transform, reordering scan after quantization and entropy coding.

A static point cloud is a frame. A slice-partitioning scheme is devised to segment a point cloud to several slices. Based on the slice level, the kd-tree method is adopted to divide the slice into macroblocks. Leafnodes of the hierarchical kd-tree are blocks. As Figure \ref{fig:layer} shows, frames, slices, macroblocks and blocks constitute the layered structure of a point cloud. 

Based on the kd-tree hierarchical structure, blocks are numbered in the breadth traversal order. Then, block-based intra prediction is introduced to reduce redundancy among adjacency blocks. Several intra modes are provided and a mode decision scheme is designed to choose the best mode. Adaptive transform tool support optimized GFT as well as DCT modes, and Lagrangian based methods are devised to mode decision. The combination of GFT and DCT improves transform efficiency than every single one mode and it can achieve better coding performance for different types of point clouds. After uniform quantization, point cloud coefficients need to be scanned to generate a one-dimensional data stream. Multiple reordering scan modes are introduced to improve the coding efficiency of entropy coding.

\section{Point Cloud Layered Structure}

A layered data structure is a fundamental representation of traditional video sequences. A video frame is divided into several slices which are independently encoded and each slice is flexibly partitioned into macroblocks. It can enhance the robustness of video encoder and improve the coding efficiency. 

Inspired by the layered structure of video coding, we devise a scheme to generates suitable layered structure in point clouds. As Figure \ref{fig:layer} shows, the scheme segments a point cloud frame into slices, macroblocks and blocks of three layers. The continuity of color attributes among adjacency points is the key to reducing redundancy. The layered structure generation scheme is adopted to cluster points with similar color attributes into a slice or a block. This structure is mirrored in the coded point cloud attribute bitstream.

Regarding the slice-partition scheme, we propose to estimate the color continuity and separate color non-smooth areas from current point cloud into slices. First, we adopt a general kd-tree method to get coding blocks and use color variances of each block to represent the color continuity. Different thresholds are set to rank the continuity. Then, we follow the rank to spilt several slices for the complex point cloud. In this paper, we use a two-slice partition scheme as an example. If the variance in a block is larger than $threshold\_1$, we regard the block as a non-smooth block, and if the proportion of non-smooth blocks among all point cloud blocks is larger than $threshold\_2$, we separate points with non-smooth color from all blocks as a slice, and the leftovers as another slice. Examples of the slice partition on two point clouds are presented in Figure\ref{fig:slice}. The $threshold\_1$ and $threshold\_2$ are off-line trained based on the statistics of color attribute coding. For a more complex point cloud, different thresholds are set and several slice partitions are supported. 

Kd-tree is a type of binary tree representing a hierarchical subdivision for a k-dimensional space. As Figure \ref{fig:overview} shows, we apply kd-tree partition on point cloud slices to get macroblocks and blocks. The essence of the kd-tree scheme in this paper is the geometry-adaptive uniform segmentation on point cloud slices. While building each hierarchy for the kd-tree, the choice of the dimension to split and the splitting points are two major factors affecting the data structure \cite{kd}.  From the x, y, z three coordinate axes, we choose the splitting dimension with the largest geometry variance, which is regarded as the principal distribution direction of points. Along the principal direction, points are in the more discrete distribution and have weaker geometry correlations among adjacent points. The midpoint in the splitting dimension is set as the splitting point, so that the two parts divided will have almost the same number of points. Loop partition stops when the division times reach the determined kd-tree depth $n$ and there will be $2^n$ leaf nodes. The process of kd-tree partition is shown in Figure \ref{fig:kdtree}.

After tree partition, we regard leaf nodes as blocks and set indexes for these blocks following a breadth-first tree traversal on the kd-tree. The indexes are the references of block coding order. The process of kd-tree partition and the index of blocks are shown in Figure \ref{fig:kdtree}. There will be $2^n$ coding blocks with the index 1, 2, ..., $2^n$.

Considering the characteristics of the human vision system, original point colors in blocks are transformed from RGB color space to YUV color space, according to the standard file ITU-R Rec. BT. 709.

\section{Block-based Intra Prediction}

The purpose of intra prediction is to make full use of the correlation among adjacency points to reduce the redundancy on color attributes. As the Table 1 shows, the number of points in point cloud blocks may be different. Points in each block are discretely and irregularly distributed in 3D space. Therefore, it is difficult to directly adopt intra prediction scheme from traditional video coding in the compression of point cloud attributes. 

After the geometry-adaptive kd-tree partition, we get a serial of numbered blocks. Based on the block structure, these blocks are regarded as one row of frame coding units with a certain order. The color average of contained points in each block represents color attributes of the block, serving as the reference of next block attribute prediction.

Multiple angular intra prediction modes referring to forward blocks and macroblocks are implemented on this serial of numbered coding blocks. Mode decision is devised to choose the best prediction mode for each block.

\begin{figure}[t]
	\centering
	\includegraphics[width=0.45\textwidth]{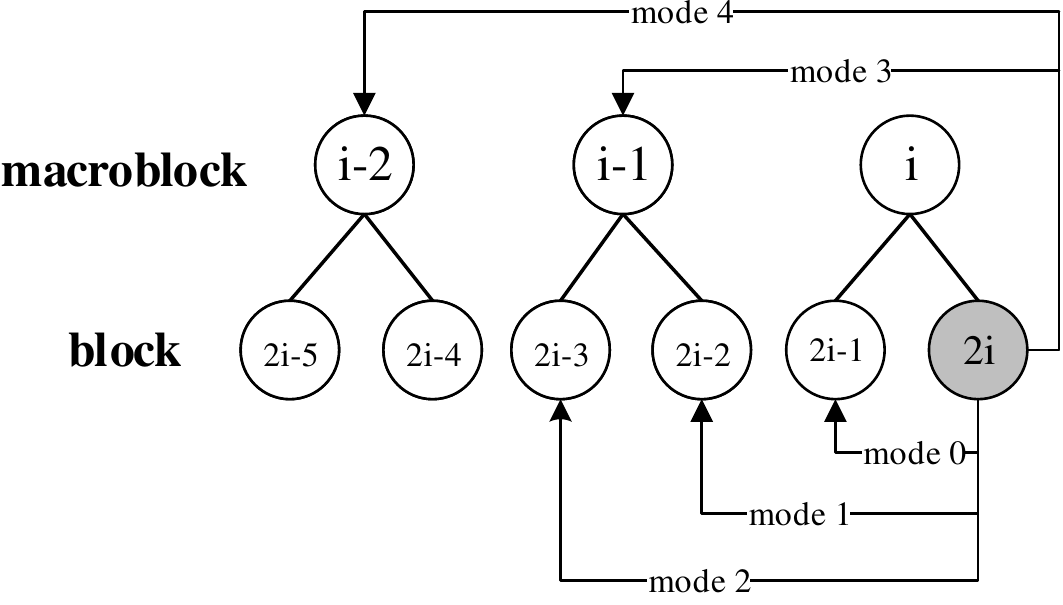}
	\caption{Multiple angular intra prediction modes for the $block_{(2i)}$ based on the tree structure.}
	\label{fig:intra}	
\end{figure}
\subsection{Several Intra Prediction Modes}

We propose multiple intra prediction modes on the blocks: DC prediction mode, three angular modes referring to three forward blocks and two angular modes referring to two forward macroblocks. If three forward blocks and two forward macroblocks are available, six prediction modes are adopted to reduce spatial redundancy on block's color attributes.

An example of block intra prediction process is presented in Figure \ref{fig:intra}, where the black circle represents current $block_{(2i)}$. Its five angular prediction references are $block_{(2i-1)}$, $block_{(2i-2)}$, $block_{(2i-3)}$, $macroblock_{(i-1)}$ and $macroblock_{(i-2)}$.

Regarding mode 5, DC prediction mode uses fixed values as the reference to Y, U, V components for current block. The first coding block must adopt DC mode.

Mode 0, mode 1 and mode 2 are three angular modes on the reference of three forward blocks. On the layer of blocks, it is an effective way to reduce the redundancy on color attributes by exploiting the correlation of adjacency spatial position.

Mode 3 and mode 4 are two angular modes referring to two forward macroblocks. A macroblock is the parent node for two child nodes. Our intra prediction framework not only supports prediction among adjacency tree leadnodes, but can also support the "parent-child" prediction scheme. The flexible prediction scheme is beneficial for attribute compression on different types of point clouds.

Each block adopts these prediction modes available to obtain residuals. The prediction residual $b_{i(res)}$ of block $b_{i}$ is $:$
\begin{small}
	\begin{eqnarray}
	b_{i(res)}=(Y_{i}-Y_{i}\_ref)+(U_{i}-U_{i}\_ref)+(V_{i}-V_{i}\_ref),
	\end{eqnarray}
\end{small}
where $Y_{i}\_ref$, $U_{i}\_ref$ and $V_{i}\_ref$ are the prediction references for current block color attributes $Y_{i}$, $U_{i}$ and $V_{i}$.

Prediction residuals are adopted in mode decision to get the best intra prediction mode.

\subsection{Prediction Mode Decision}

In rate-distortion optimization (RDO) mode selection on traditional video coding, the real encoding rate and distortion value usually need to be calculated by performing transform, quantization, entropy coding, inverse quantization and inverse transform. The process is time-consuming and computationally complex. 

Inspired by the fast intra mode decision method for H.264/AVC in ~\cite{SATD}, in order to reduce the computation cost and maintain coding performance, we develop SATD as the cost criteria for point cloud intra mode decision.   

SATD represents the sum of the absolute transformed differences between current block's attributes and the reference values, which is a comprehensive consideration of rate and distortion. In our paper, after intra prediction on the current block, block attribute residuals are transformed by DCT. Then SATD is calculated to estimate the prediction performance, which is presented as $:$
\begin{eqnarray}
SATD = sum(abs(DCT(b_{i(res)}))).
\end{eqnarray}

The smaller the SATD, the better the prediction performs. The intra prediction mode with the smallest SATD will be chosen as the best intra prediction mode.

\section{Adaptive Transform}
\subsection{Two transform modes}

Graphs have flexible geometric structures, which are the natural representations of 3D irregular point clouds. Composed of vertex and edges, graphs preserve more underlying information about the real 3D structure and the correlations among points. Current graph transform scheme in \cite{shao} uses kd-tree partition to get coding blocks and adopts Laplacian sparsity optimized GFT in each block, which is demonstrated to achieve better transform efficiency than other works. In this paper, we adopt optimized GFT as one of transform modes.

For each coding block, we form a graph by connecting all points with edges. Define the graph $G$ as :
\begin{eqnarray}
G = (v = ({ n_{1},n_{2},...,n_{i})},\xi ).
\end{eqnarray}
where $n_{i}$ represents the node in the graph $G$ and $\xi$ represents the sets of edges.

In the adjacency matrix $W$, the edge weight $w_{i,j}$ describes the correlation between two nodes $n_{i}$ and $n_{j}$ by the geometry distance. $w_{i,j}$ is presented as: 
\begin{eqnarray}
w_{i,j}=\left\{\begin{matrix} e^-{\frac{(n_{i}-n_{j})^2}{\delta^2 }},& if (n_{i}-n_{j})^2\leq \tau;  \\ \\ 
0, & else, 
\end{matrix}\right.
\end{eqnarray}
where $\delta$ denotes the variance of graph nodes and $\tau$ is the Euclidean distance threshold between two nodes.

The degree matrix $D$ is a diagonal matrix indicating the popular degree of points. $D_{i}$ is illustrated as:  
\begin{eqnarray}
D_{i}=\sum_{j}{w_{i,j}}.
\end{eqnarray}

We choose the Laplacian matrix $L$ presented in Equation (6) as the graph shift operator.
\begin{eqnarray}
L=D-W.
\end{eqnarray}

The graph transform matix $A$ in Equation (7) is the eigenvector matrix of the Laplacian matrix $L$.
\begin{eqnarray}
L=A\lambda A^{-1},
\end{eqnarray}
where $\lambda$ is a diagonal matrix with eigenvalues of $L$.

Then the graph transform matrix $A$ is used to decorrelate point cloud attributes from spatial domain to graph spectrum domain.

However, GFT is not suitable for all types of point clouds. For example, for sparse point clouds, Euclidean distances among points are generally large so it is difficult to construct an effective graph by estimating the underlying relationship between relative geometry position and color attributes. An efficient graph transform matrix cannot be derived under this circumstance. Regarding coding a kind of point cloud with complex geometry but flat color attributes, the GFT scheme cannot work well on attribute compression because it is too dependent on point cloud geometry information. But DCT scheme can handle those kinds of point clouds well. 

General one-dimensional DCT is the other transform mode in our transform scheme, which is combined with optimized GFT to deal with different types of point clouds.

\begin{figure}[t]
	\centering
	\includegraphics[width=0.51\textwidth]{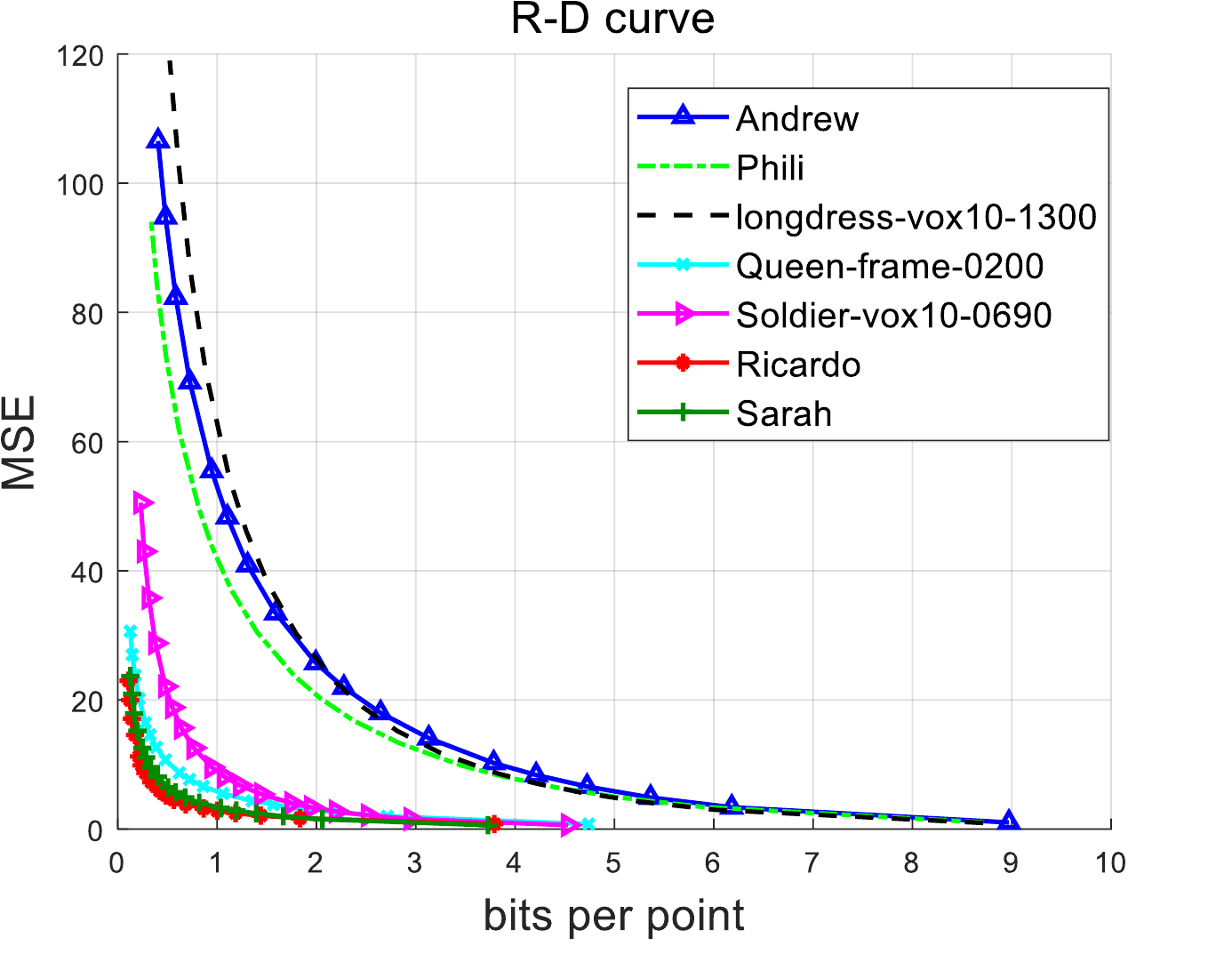}
	\caption{Fitted RD curves on several operating points for point cloud test sets.}
	\label{fig:mse}	
\end{figure}

\begin{figure}[t]
	\centering
	\includegraphics[width=0.5\textwidth]{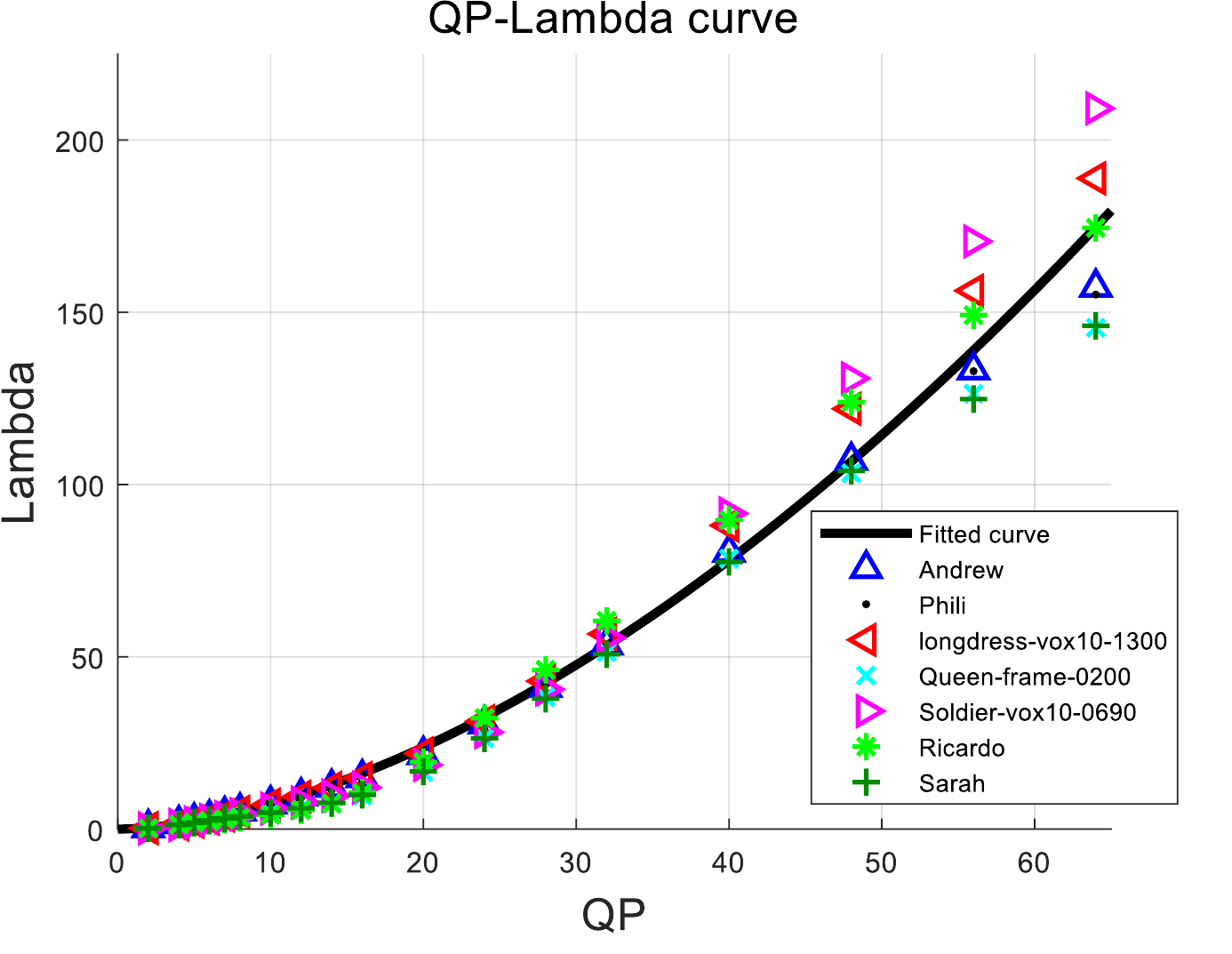}
	\caption{The relationship between quantization step $Q$ and Lagrange multiplier $\lambda$}
	\label{fig:lambda}	
\end{figure}

\subsection{Lagrangian-based Mode Decision}

To achieve better transform efficiency for different point clouds, it is necessary to devise a mode decision scheme to adaptively select the best transform mode. Different transform modes have different rate-distortion characteristics and the goal of the mode decision is to optimize its overall performance and reach the best balance between rate and distortion. That is, minimize distortion $D$, subject to a constraint $R_{c}$ on the number of bits used $R$ \cite{rdo}. The constrained problem is illustrated as $:$

\begin{eqnarray}
min\ D\ \ \ \ subject\ to\ R < R_{c}.
\end{eqnarray}

In traditional video coding, Lagrange multiplier optimization technique is usually adopted to solve the optimization task in Equation (8). The Lagrangian formulation of the minimization problem is given:

\begin{eqnarray}
min\ J\ \ \ \ where\ \ J = D + \lambda R,
\end{eqnarray}
where $J$ is the RD cost of a certain mode and $\lambda$ is the Lagrange multiplier to balance the trade-off between rate $R$ and distortion $D$. The mode with the minimum $J$ will be chosen as the best mode.

Before we calculate RD cost $J$ in function (9), bitrate and distortion of each mode need to be estimated and the Lagrange multiplier $\lambda$ need to be determined.

\subsubsection{Rate and Distortion Estimation}

The bitrate $R$ of a transform mode is calculated in terms of the bits per point (bpp) of point cloud attribute transformed coefficients after quantization and entropy coding. Meanwhile, more bits are costed for signaling the mode information.

The distortion $D$ of a transform mode is estimated by the mean square error (MSE) between YUV components before transform and the reconstructed YUV components after the processing of transform, quantization, inverse quantization and inverse transform shown in Figure \ref{fig:overview}.

\subsubsection{Lagrangian Multiplier Derivation based on $\lambda$-Q model}

In traditional video coding, a Lagrangian optimization method is a well-established scheme to mode decision and the Lagrange multiplier $\lambda$ can be off-line trained based on a mathematical function of quantization step $Q$. 

In this paper, we refer to a point cloud Lagrangian optimization method in \cite{xu} to derive Lagrange multiplier $\lambda$ for our proposed attribute compression scheme.

The goal of transform mode decision is to find the mode with the minimum RD cost $J$ in Function (9). To solve the optimization task, we take the derivative of $J$ to quantization step $Q$ in Function (10).

\begin{eqnarray}
\frac{\partial J}{\partial Q} = \frac{\partial D}{\partial Q} + \lambda \frac{\partial R}{\partial Q}.
\end{eqnarray}

When the derivative $\frac{\partial J}{\partial Q}$ equals 0, the Lagrange multiplier $\lambda$ is derived in Function (11) and the minimum RD cost $J$ is achieved. It implies that $\lambda$ is the slope of the RD performance.

\begin{eqnarray}
\lambda = -\frac{\partial D}{\partial R}.
\end{eqnarray}

We choose some typical point cloud datasets as test sets for our compression scheme and record rate and distortion values at several different $Q$. Fitted RD curves based on those rate and distortion operating points for point cloud test sets are presented in Figure \ref{fig:mse}. Then we use the slope of the rate and distortion differences between current operating point and neighboring operating points to approximate the Lagrange $\lambda$ in Function (11). The mathematical relationship between the Lagrange multiplier $\lambda$ and quantization step $Q$ is estimated based on the statistics of point cloud color coding performance. 

Therefore, the $\lambda-Q$ model for our proposed compression scheme can be approximated as:

\begin{eqnarray}
\lambda = a * Q^{^{b}},
\end{eqnarray}
where $a$ = 0.14 and $b$ = 1.72.

In statistics, the coefficient of determination $R-square$ usually serves as the measure of the fitting degree. In our trained $\lambda-Q$ model, $R-square$ is 0.98, which it implies $\lambda$ and $Q$ are well fitted on the fitting curve in Figure \ref{fig:lambda}. 

Based on the trained $\lambda-Q$ model, Lagrangian optimization can be achieved to adaptively select the best transform mode for different blocks of the point cloud.

\begin{figure}[t]
	\centering
	\includegraphics[width=0.4\textwidth]{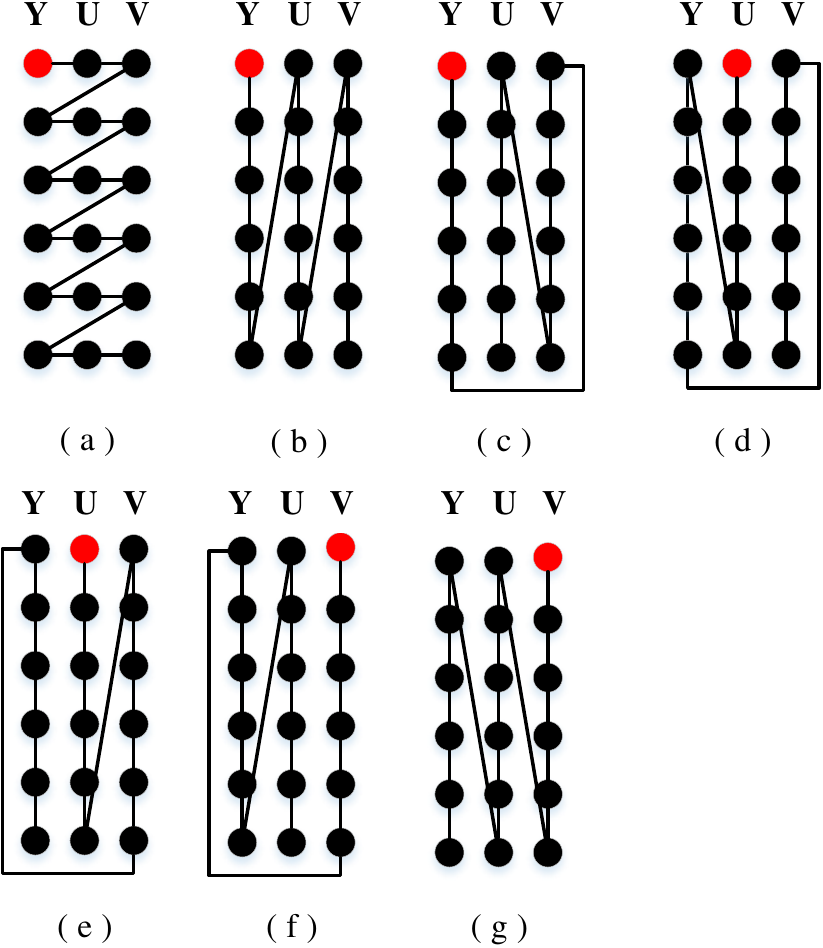}
	\caption{Multiple reordering scan modes for block attributes before entropy coding: (a) mode 0; (b) mode 1; (c) mode 2; (d) mode 3; (e) mode 4; (f) mode 5; (g) mode 6.}
	\label{fig:scan}	
\end{figure}

\section{Reordering Scan of Quantized Coefficients}

After transform and uniform quantization, YUV component coefficients of all points in a point cloud are processed. For a block, color information is presented as a $n$$\times$$3$ coefficient matrix. Each element in the matrix is a luminance or chrominance component for a certain point. Before entropy coding, in a block unit, all the YUV component coefficients need to be scanned and reordered into a string of coefficients stream. 

In order to improve entropy coding efficiency, to increase the length of continuous zero component, seven reordering scan modes are supported for every quantized coefficients block. The schematic diagram for seven scan modes is presented in Figure \ref{fig:scan}. The three elements in each row are Y component, U component and V component of a certain point, respectively. The red point is the beginning of the scan process. 

Mode 0 is a raster scan mode. It horizontally scans the first point's YUV component and then scans the next point's components . Mode 1 to mode 6 vertically scan a certain kind of component in all points and follow the specified path to scan other components. The differences among mode 1 to mode 6 are the scan beginning and the scan order for YUV components. 

After the reordering scan process, the continuous zero sequence in the bottom of bitstream is discarded to cut the unnecessary bit expenses. The reordering scan mode with the longest continuous zero in the bottom of bitstream is the selected scan mode.

\section{Experimental Results}

To compare the performance of the proposed method with the state-of-the-art RAHT scheme \cite{ricardo} and the TMC1 anchor latest released in MPEG 121st meeting, we have conducted many tests using different point cloud frames. According to the MPEG PCC PSNR evaluation proposal \cite{psnr} and the BD-BR performance evaluation scheme on traditional video coding \cite{bdbr}, our method achieves significant coding performance improvement compared with the RAHT scheme and the MPEG TMC1 anchor.

\subsection{Datasets}

\begin{figure}[t]
	\centering
	\includegraphics[width=0.51\textwidth]{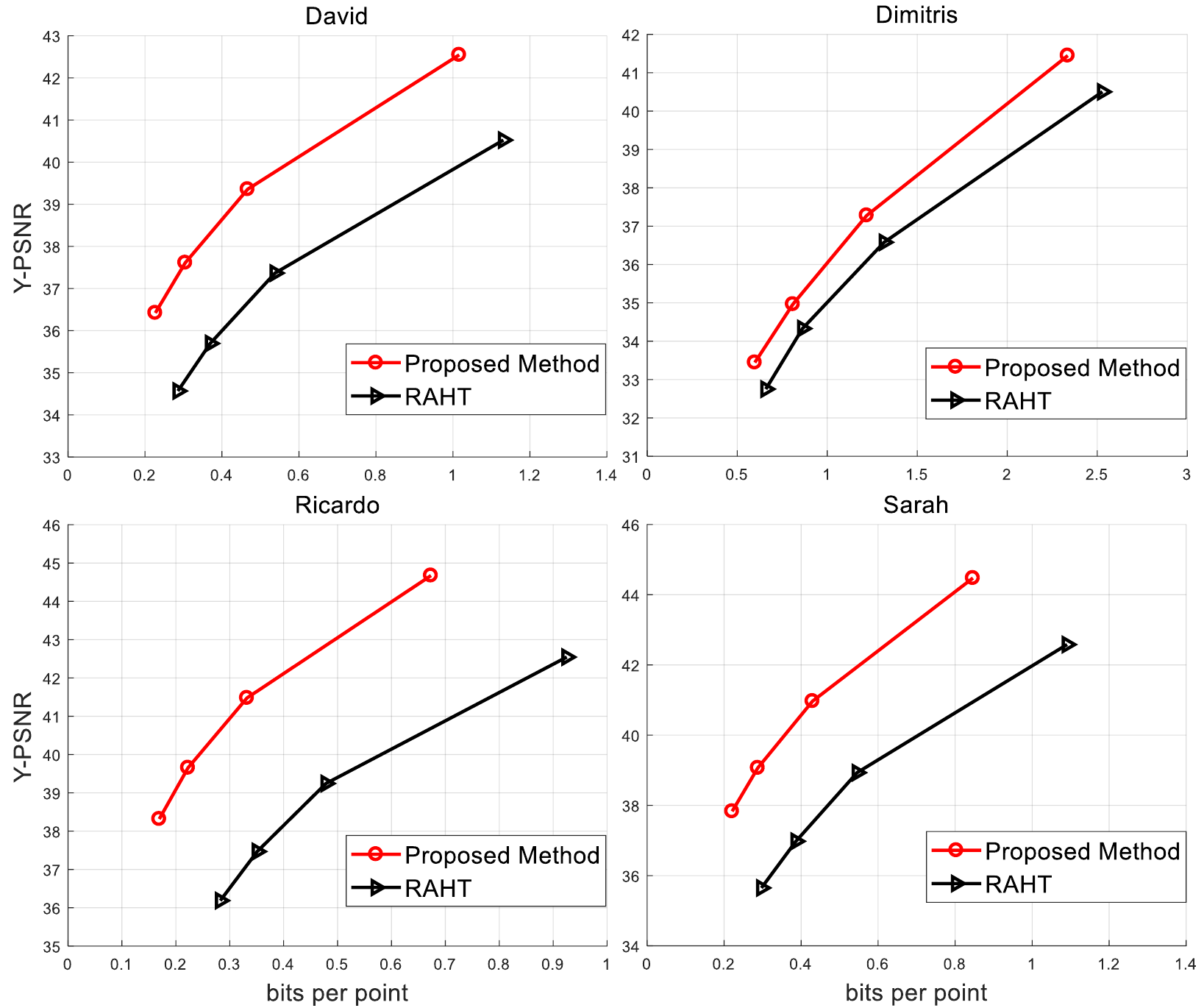}
	\caption{Comparison of coding performance between the proposed method and the RAHT scheme.}
	\label{fig:raht}	
\end{figure}

\begin{table}
	\caption{BD-rate of the Proposed Method and the RAHT Scheme}
	\label{tab:raht}
	\footnotesize
	\begin{tabular}{ccccl}
		\hline
		Point Cloud&BDBR$-$Y&BDBR$-$U&BDBR$-$V&BDBR$(AVG)$\\
		\hline
		Boy & -61.12\% & -65.0\%& -65.0\%&\ \ \ \ \ -63.72\%\\
		David & -46.04\% & -23.2\%& -19.2 \%&\ \ \ \ \ -29.48\%\\
		Andrew & -19.19\% & 0.8\%& 2.9\%&\ \ \ \ \ -5.16\%\\
		Phili & -17.69\% & 6.7\%& 15.4\%&\ \ \ \ \ 1.46\%\\
		Ricardo & -56.18\% & -44.6\%& -37.5\%&\ \ \ \ \ -46.09\%\\
		Sarah & -47.37\% & -30.4\%& -25.7\%&\ \ \ \ \ -34.48\%\\
		Dimitris & -18.05\% & -32.1\%& -34.3\%&\ \ \ \ \ -28.15\%\\
		\hline
		Average & -37.95\% & -26.83\%& -23.34\%&\ \ \ \ \ -29.37\%\\
		\hline
	\end{tabular}
\end{table}  
While comparing with the RAHT scheme, we adopt $Ricardo$, $Sarah$, $Andrew$, $Phili$ four datasets tested in \cite{ricardo} and $Boy$, $David$, $Dimitris$ three other datasets. As for comparing with the MPEG TMC1 anchor, we follow the common test conditions (CTC) for category 1.2 in \cite{ctc}. We use $Facade15\_vox14$, $Facade15\_vox20$, $Queen\_frame\_0200$ and $Soldier\_vox10\_0690$ of class A and $Shiva35\_vox12$, $Shiva35\_vox20$, $Longdress\_vox10\_0300$ of class B.

\begin{figure}[t]
	\centering
	\includegraphics[width=0.47\textwidth]{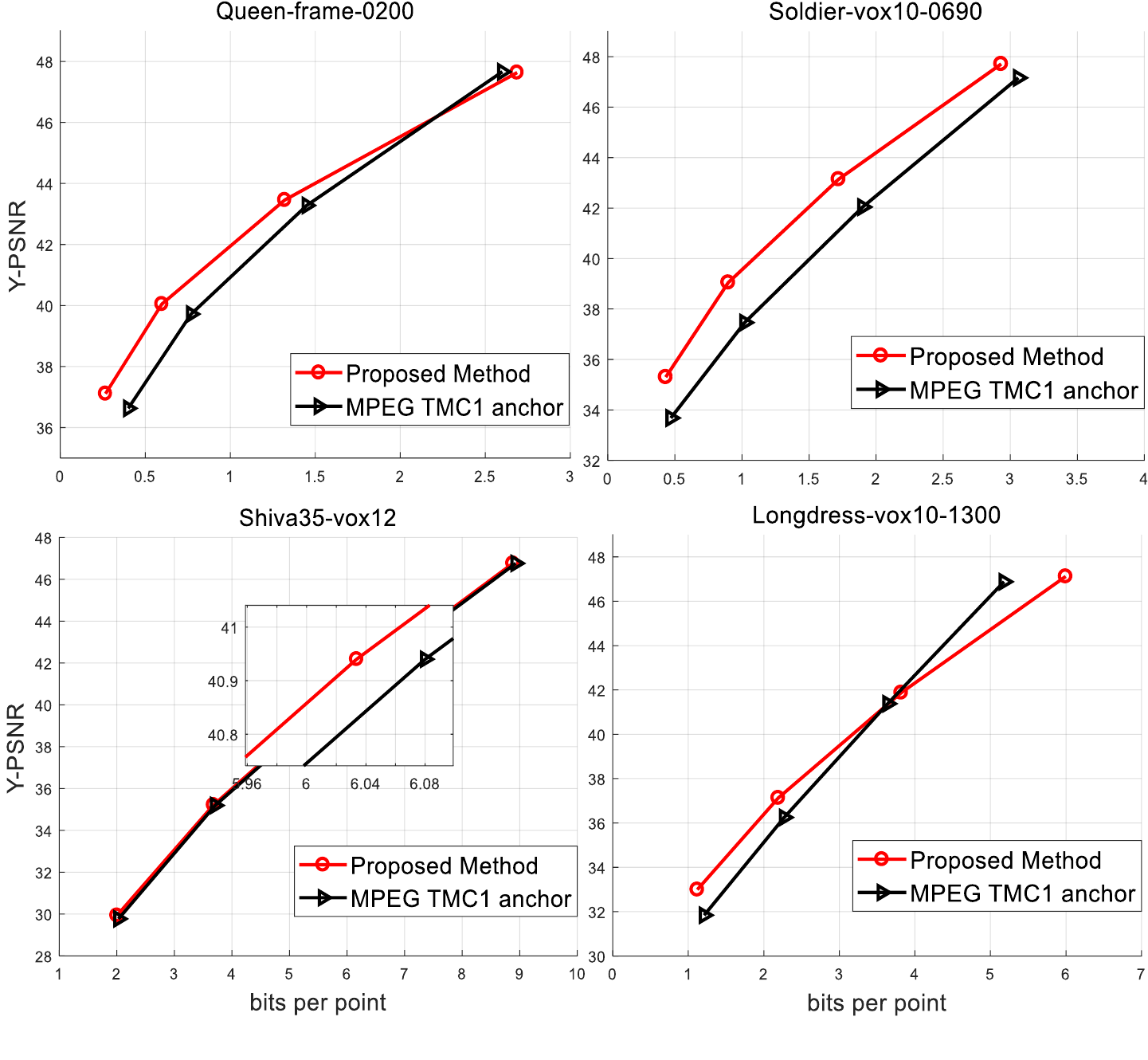}
	\caption{Comparison of coding performance between the proposed method and the TMC1 anchor.}
	\label{fig:mpeg}	
\end{figure}

\begin{table}
	\caption{BD-rate of the Proposed Method and the MPEG TMC1 Anchor}
	\label{tab:freq}
	\footnotesize
	\begin{tabular}{ccccl}
		\hline
		Point Cloud&BDBR$-$Y&BDBR$-$U&BDBR$-$V&BDBR$(AVG)$\\
		\hline
		Queen\_frame\_0200 & -17.96\% & -62.4\%& -68.4 \%&\ \ \ \ -49.58\%\\
		Soldier\_vox10\_0690 & -25.11\% & -22.0\%& -21.1\%&\ \ \ \ -22.74\%\\
		Shiva35\_vox12 & -1.16\% & -4.9\%& -5.0\%&\ \ \ \ -3.69\%\\
		Shiva35\_vox20 & -1.61\% & -4.0\%& -4.6\%&\ \ \ \ -3.41\%\\
		Longdress\_vox10\_1300 & -5.15\% & -40.1\%& -55.8\%&\ \ \ \ -33.68\%\\
		Facade\_00015\_vox14 & 7.02\% & -6.7\%& -2.0 \%&\ \ \ \ -0.57\%\\
		Facade\_00015\_vox20 & 6.39\% & -6.9\%& -2.3 \%&\ \ \ \ -0.94\%\\
		\hline
		Average & -5.37\% & -21.01\%& -22.74\%&\ \ \ \ -16.37\%\\
		\hline
	\end{tabular}
\end{table} 

\subsection{Implementation details}
For kd-tree partition, considering the compression performance and the computation complexity, the number of points in each coding block is limited to the empirical range (100, 200). The details of kd-tree partition on two point clouds are shown in Table 1.

About the training of $\lambda-Q$ model, we use $Andrew$, $Phili$, $Ricardo$, $Sarah$ of the RAHT work and $Queen\_frame\_0200$, $Soldier\_vox104\_0690$ of MPEG class A and $Longdress\_vox10\_0300$ of MPEG class B to estimate the Lagrange multiplier $\lambda$. 

We adopt uniform quantization and arithmetic entropy coding on transformed attribute residuals. Intra prediction modes, transform modes, scan modes, the number of discarded zero for each block and block attribute residuals are encoded into bitstreams. Different quantization steps are used to reach the test bitrate point and to obtain different pairs of bitrate and PSNR. We use the bits per point (bpp) to measure the total bitrate of Y, U, V three components for each point. PSNR is calculated on the Y component by the evaluation metric from MPEG PCC standard proposal \cite{psnr}.

\subsection{Objective coding performance evaluation}

The performance comparisons between the proposed method and the RAHT scheme are tabulated in Table 2. The results in Table 2 show that the proposed method obtains a 37.95\% BD-rate gain in luma component, a 26.83\% and a 23.34\% BD-rate gain for two chroma components, respectively. On average, a 29.37$\%$ BD-rate gain can be achieved. Moreover, the rate-distortion performance comparisons for point cloud $David$, $Dimitris$, $Ricardo$ and $Sarah$ are shown in Figure \ref{fig:raht}. Our proposed method achieves significant coding performance improvement compared with the RAHT scheme on those datasets and the PSNR gain for Y component can be up to 4 dB.

The performance comparisons between the proposed method and the MPEG TMC1 anchor are tabulated in Table 3. The results in Table 3 show that the proposed method achieves a 5.37\% BD-rate gain in luma component, a 21.01\% and a 22.74\% BD-rate gain for two chroma components, respectively. On average, a 16.37$\%$ BD-rate gain can be achieved. Moreover, the rate-distortion performance comparisons for four point cloud contents $Queen\_frame\_0200$, $Soldier\_vox10\_0690$, $Shiva35\_vox12$ and $Longdress\_vox10\_0300$ are shown in Figure \ref{fig:mpeg}. Because of the well-established intra prediction scheme, our proposed method achieves better performance than the TMC1 for low bit rate coding. On point cloud $Soldier\_vox10\\ \_0690$, the coding performance at high bit rate coding is also better than TMC1 anchor. Regarding the performance at high bit rate coding on other three datasets, there are some improvement need to be done on our transform scheme in future. Overall, our proposed method performs better than the TMC1 anchor in Table 3.

\subsection{Ablation study}

To further analyze the contributions of different coding tools in our proposed scheme, namely slice-partition, block-based intra prediction, adaptive transform and reordering scan, for point cloud attribute compression, we implement an ablation study.

We construct four models, which are described as follows. First model V1 in \cite{psnr} is regarded as the baseline of our work. It adopts the kd-tree and optimized GFT to compress point cloud attributes. The second model V2 adds an adaptive transform tool to the baseline and third model V3 implements the intra prediction tool to the V2. The slice partition scheme is adopted in V4 model. On the basis of model V4, we introduce the reordering scan method to process the bitstream, which completes our full framework on point cloud attribute compression.

\begin{figure}[t]
	\centering
	\includegraphics[width=0.54\textwidth]{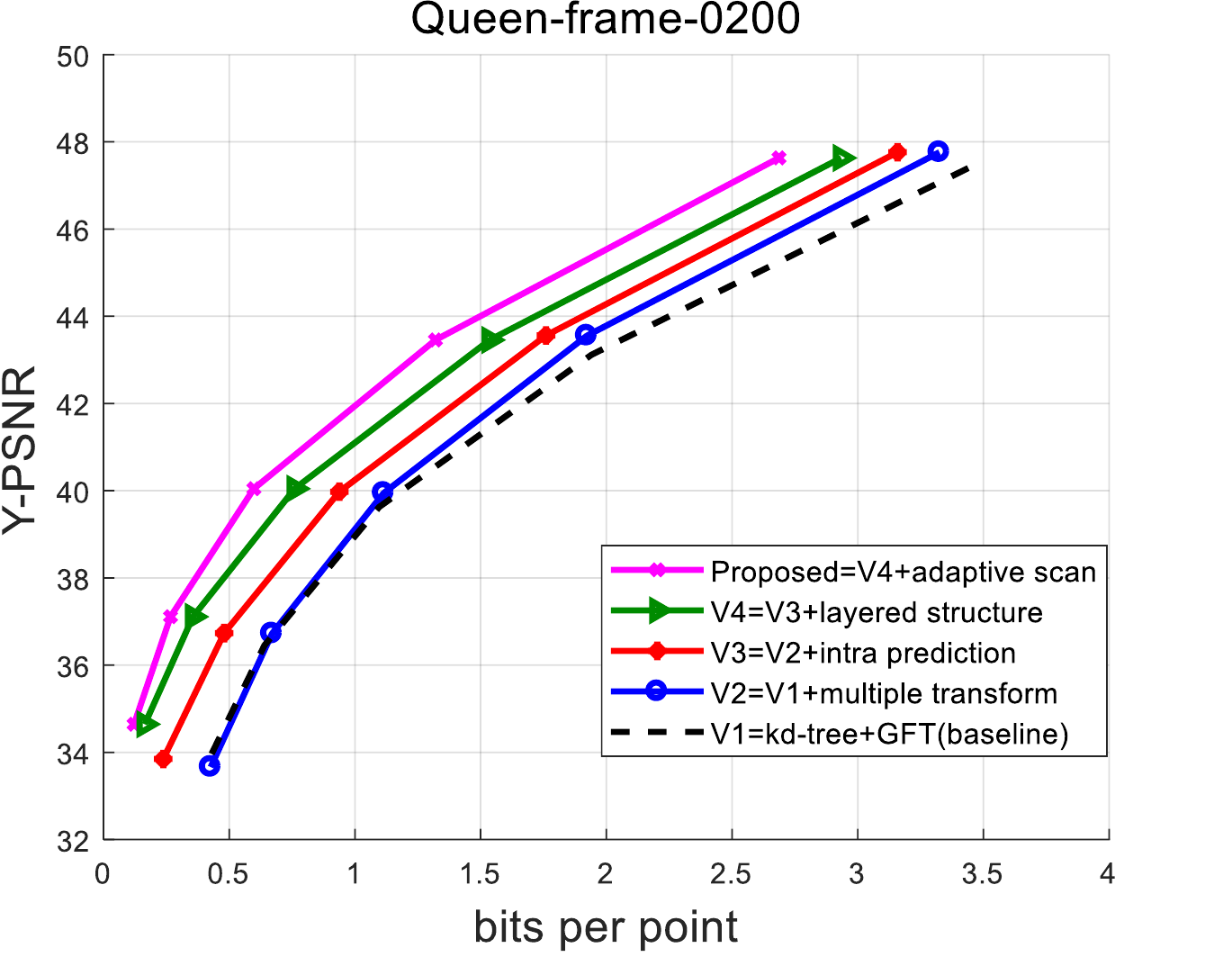}
	\caption{An ablation study of proposed point cloud attribute compression scheme.}
	\label{fig:inside}	
\end{figure}

The comparison of coding performance among five models on point cloud $Queen\_frame\_0200$ is presented in Figure \ref{fig:inside}.Experimental results show that our four coding tools all bring good gains to the R-D performance. 

\section{Conclusions}

In this paper, we propose an efficient hybrid point cloud attribute compression scheme. The novelty of the proposed scheme lies in layered structure generation and block-based intra prediction. Moreover, the adaptive GFT-based transform is Lagrangian optimized and the Lagrange multiplier is off-line derived based on the statistics of color attribute coding. Multiple reordering scan modes are dedicated to improve coding efficiency for entropy coding. Experimental results demonstrate that our method performs significantly better than the state-of-the-art RAHT system, and on average a 29.37$\%$ BD-rate gain is achieved. Comparing with the TMC1 anchor's coding results on MPEG 121st meeting, on average a 16.37$\%$ BD-rate gain is obtained. 

\bibliographystyle{IEEEbib}
\bibliography{icme2018template}

\end{document}